%% file: main.tex
\documentclass[conference]{IEEEtran}
\IEEEoverridecommandlockouts
\usepackage{cite}
\usepackage{amsmath,amssymb,amsfonts}
\usepackage{algorithmic}
\usepackage{graphicx}
\usepackage{textcomp}
\usepackage{xcolor}
\usepackage{subcaption}

\usepackage{gensymb}
\usepackage{bbm}

\usepackage{pgfplots}

\usepackage{pgfplots}
\usepackage{svg}  
\usetikzlibrary{plotmarks}
\usetikzlibrary{arrows.meta}
\usepgfplotslibrary{patchplots}

\def\BibTeX{{\rm B\kern-.05em{\sc i\kern-.025em b}\kern-.08em
    T\kern-.1667em\lower.7ex\hbox{E}\kern-.125emX}}
\begin{document}

\title{
Statistical Framework for Clustering  MU-MIMO Wireless via Second Order Statistics
\thanks{
The work of Roberto Pereira has been funded by the grant  CHIST-ERA-20-SICT-004 (SONATA) by PCI2021-122043-2A/AEI/10.13039/501100011033 and the work of  Xavier Mestre was supported by the grant PID2021-128373OB-I00 funded by MCIN/AEI/10.13039/501100011033 and by “ERDF A way of making Europe”. 
}
}

\author{\IEEEauthorblockN{Roberto Pereira and Xavier Mestre}
\IEEEauthorblockA{Centre Tecnol\`ogic de Telecomunicacions de Catalunya
(CTTC/CERCA)\\
Av. Carl Friedrich Gauss 7, 08860 Castelldefels, Spain \\
email: \{rpereira, xmestre\}@cttc.es}
}

\newtheorem{thm}{Theorem}[section]
\newtheorem{lem}[thm]{Lemma}
\newtheorem{prop}[thm]{Proposition}
\newtheorem{cor}{Corollary}
\newtheorem{conj}{Conjecture}[section]
\newtheorem{defn}{Definition}[section]
\newtheorem{exmp}{Example}[section]
\newtheorem{rem}{Remark}

\newtheorem{theorem}{Theorem}[section]
\renewcommand{\IEEEQED}{\IEEEQEDopen}

\maketitle


\begin{abstract}

This work explores the clustering of wireless users by examining the distances between their channel covariance matrices, which reside on the Riemannian manifold of positive definite matrices.
Specifically, we consider an estimator of the Log-Euclidean distance between multiple sample covariance matrices (SCMs) consistent when the number of samples and the observation size grow unbounded at the same rate. 
Within the context of multi-user MIMO (MU-MIMO) wireless communication systems, we develop a statistical framework that allows to  accurate predictions of the clustering algorithm's performance under realistic conditions.
Specifically, we present a central limit theorem that establishes the asymptotic Gaussianity of the consistent estimator of the log-Euclidean distance computed over two sample covariance matrices.
\end{abstract}

\begin{IEEEkeywords}
Clustering, Wireless Communication, Multiple-Antenna Systems, Central Limit Theory.
\end{IEEEkeywords}

\section{Introduction}

Modern wireless communication systems rely on multiple-antenna radio access technologies to enhance the overall connectivity and spectral efficiency  of the network. The use of multi-antenna technology allows to introduce an additional orthogonality dimension that can be combined with time and/or frequency multiplexing to enhance the total achievable rates in a wireless setting.
When the total number of user equipments (UE) is larger than the number of antennas at the base station (BS), spatial multiplexing alone is not able to completely cancel out interference among transmissions. In this type of setting,  several advanced non-orthogonal techniques have been proposed in the literature, which try to manage this residual interference. Examples of this type of approach are non-orthogonal multiple access (NOMA) \cite{saideh2019joint_noma}, joint spatial division multiplexing (JSDM) \cite{adhikary2013joint_spatial_first} and  rate splitting (RSMA) \cite{rs_2rhs} to name a few.
All these methods need an effective partitioning of receivers into clusters according to spatial proximity, so that specific signal processing is applied to separately treat inter-cluster and intra-cluster interference. 

Typically, one may use spatial precoding  to separate clusters of users in the angle domain. In this case, the similarity among principal channel subspaces can be taken as a relevant measure of angle proximity~\cite{roberto21_globecom, flores2023clustered, agglomerative_user_clustering}. One may therefore use orthogonal multiplexing to separate users in the same angular cluster. However, this is not enough in heavily loaded systems, where one may additionally exploit the power domain to separate users within the same angular characteristics. The idea here is that simultaneous transmission of signals with very different power may be possible with the use of sequential interference cancellation~\cite{adhikary2013joint_spatial_first, clerckx2021noma_rate_splitting}
so that one may
cluster users according to their spatial proximity (both in terms of their angles of arrival and distance to the BS). Inter-cluster separability can be attempted by spatial multiplexing and sequential interference cancellation, whereas intra-cluster separability can be achieved by orthogonal techniques. This previously requires clustering the channels of the different UEs according to their spatial proximity. Since wireless channels experiment rapid fluctuations while multiplexing schemes need to be fixed for long periods of time, it seems convenient to carry out this clustering by relying on the second order statistics of the channels, rather than the channel realizations themselves. In order to perform this clustering, it seems convenient to rely on distances between spatial covariance matrices. 

Recent approaches have shifted focus from the traditional Euclidean distance\cite{li2013riemannian, roberto_icaspp2023consistent} to the study of distances that exploit the fact that covariance matrices belong to the Riemmannian manifold of positive definite matrices \cite{lhuang2017riemannian,shinohara2010covariance,li2013riemannian_euclidean,Barachant13,shi2019riemannian}. 
This is the case of the (square) log-Euclidean distance \cite{arsigny06} which, given two covariance matrices $\mathbf{R}_1$, $\mathbf{R}_2$, is defined as 
\begin{equation} 
\label{eq:defLE}
{d}_{M}(1,2) =\frac{1}{M}\mathrm{tr}\left[  \left(  \log\mathbf{R}_{1}%
-\log\mathbf{R}_{2}\right)  ^{2}\right]
\end{equation}
where  the logarithm is applied matrix-wise (i.e. to the eigenvalues).
The log-Euclidean metric was originally derived by endowing the manifold of positive definite matrices with an appropriate Lie group structure, together with a logarithmic scalar multiplication that gives the essential properties of a vector space \cite{arsigny07}. 
Contrary  to other 
alternatives, e.g., affine-invariant metric \cite{ilea2018covariance, li2013riemannian_euclidean}, the log-Euclidean distance is more amenable from the computational complexity,  has a closed form solution for its (Fréchet) mean and always yields a positive definite Gaussian kernel~\cite{Jayasumana15}. Hence, it is often the
choice when comparing 
covariance matrices \cite{ilea2018covariance,li2013log,Jayasumana15,wang2012covariance}.

Unfortunately, in practical scenarios, the BS only has access to the instantaneous channel of each transmitter and has to estimate the distance between spatial covariance matrices using raw channel data. The idea here is that the BS can periodically estimate the channel of each UE (e.g. by the use of training symbols), and then process this information to obtain the distance between covariance matrices. The simplest way to carry out this task would be to estimate the distance between covariance matrices as the distance between \emph{sample} covariance matrices (SCM). Unfortunately, SCMs are not reliable estimates of the true ones when the number of channels realizations is not sufficiently large compared to the number of antennas at the BS. This introduces variability and uncertainty into the obtained distances, affecting the accuracy of transmitter grouping and the subsequent network communication performance. 
In these situations, it becomes crucial to ensure that the distance estimators are consistent when the number of samples per transmitter is finite and comparable to the number of antennas at the BS. In \cite{roberto_icaspp2023consistent}, we  derive a consistent estimator of the log-Euclidean metric in (\ref{eq:defLE}) which provably improves the 
comparison of two SCMs. 


This paper tries to quantify the performance of the proposed distance estimators when attempting to cluster UEs in a practical wireless communications setting. This can be done by carrying out a statistical analysis of the different distances to quantify how effective they are in a certain scenario. 
Given the mathematical complexity of the proposed distances, this analysis may typically involve some sort of Monte Carlo or bootstrap simulation mechanism~\cite{kimes2017statistical, liu2008statistical} which are often computationally expensive and require large samples sizes to obtain accurate results. Motivated by the need for an efficient statistical framework, we present here a
Central Limit Theorem (CLT) for the consistent estimator of the log-Euclidean which, in turn, allows us to characterize and predict the quality of clustering of multiple transmitter devices under realistic wireless communication settings.

\section{Preliminaries}
\label{sec:preliminaries}


Let us consider a wireless  scenario where a base station (BS) equipped with $M$ antennas simultaneously communicates to $K$ single-antenna user equipments (UEs) over an uplink channel. 
We assume that the BS has access to $N_j$ independent realizations of the $j$th user's channel matrix and denote as $\mathbf{Y}_j \in \mathbb{C}^{M \times N_j}$ the  matrix  containing these $N_j$ realizations  as columns,  for $j = 1, \dots, K$.
Moreover, we consider a correlated a Rayleigh Fading 
channel model such that the channel matrix of the $j$th user is generated as 
\begin{equation}
    \label{eq:preliminar:channel}
    \mathbf{Y}_j = \mathbf{R}_{j}^\frac{1}{2}\mathbf{X}_j
\end{equation}
where $\mathbf{X}_{j}$ is an $M \times N_j$ matrix of 
i.i.d.
entries with zero mean and unit variance associated to  the small-scale fading whereas $\mathbf{R}_{j} \in \mathbb{C}^{M\times M}$ is the channel covariance matrix at the BS related to the $j$th user.  
 Specifically, $\mathbf{R}_{j}$  
models  the channel's spatial correlation and large-scale fading $\beta_j = M^{-1}\mathrm{tr}(\mathbf{R}_j), j = 1, \ldots, K$, which depend on the angular position and distance between the UE and the BS, respectively.

\subsection{ Estimators of log-Euclidean Distance}


Following the discussion above, we are interested in clustering the different transmitters such that UEs with similar covariance matrices belong to the same cluster. Unfortunately, the BS only has access to an estimator of these covariance matrices, namely the SCM 
$$
\hat{\mathbf{R}}_{j}=\frac{1}{N_{j}}\mathbf{Y}_{j}\mathbf{Y}_{j}^\mathrm{H}, \quad j = 1, \dots, K
$$ 
associated to each of the UEs. 
In this context, for a pair of UEs $(r,s), r \neq s, r,s = 1,\ldots, K$, the BS may attempt to estimate ${d}_{M}(r,s)$  by plugging the SCMs $\hat{\mathbf{R}}_{r}, \hat{\mathbf{R}}_{s}$ into (\ref{eq:defLE}) resulting in the traditional \textit{plug-in} estimator, hereafter, defined as 
$$
\tilde{d}_{M}(r,s) =\frac{1}{M}\mathrm{tr}\left[  \left(  \log\hat{\mathbf{R}}_{r}%
-\log\hat{\mathbf{R}}_{s}\right)  ^{2}\right].
$$
For clarity, in this section, we will assume $K=2$ and therefore $r=1,s=2$. In the next section, we will explore the more general case where $K > 2$. 
Furthermore, we will omit the indices $(1,2)$ whenever it is clear from the context and write, for instance,  $d_M = d_M(1,2)$. 

As discussed above, in the wireless communications context, where the number of channel samples $N_1, N_2$ is comparable to the number of antennas at the BS, the \textit{plug-in} estimator often fails to correctly approximate $d_M$. In this setting, it is more reliable to use another form of distance estimator, which turns out to be consistent even if $N_1, N_2$ increase with $M$, see further \cite{roberto_icaspp2023consistent}. 
Let us now introduce some formal mathematical assumptions that will guarantee the consistency of the proposed estimator. 

We make the following assumptions on the covariance matrices $\mathbf{R}_j, j = 1, \ldots, K$:

\noindent \textbf{(As1)} The different eigenvalues of $\mathbf{R}_{j}$ are denoted $0<\gamma_{1}^{(j)}<\ldots<\gamma_{\bar{M}_{j}}^{(j)}$, for $j = 1, \ldots, K$, \ and have multiplicity $K_{1}^{(j)},\ldots,K_{\bar{M}_{j}}^{(j)}$, where $\bar{M}_{j}$ is the total number of distinct eigenvalues. All these quantities may vary with $M$ \ but we always have $\inf_{M}\gamma_{1}^{(j)}>0$ and $\sup_{M}
\gamma_{\bar{M}_j}^{(j)}<\infty$.

\noindent \textbf{(As2)} The quantities $N_{j}, j = 1, \ldots, K$ depend on $M$, that is
$N_{j}=N_{j}(M)$. Furthermore, when $M\rightarrow\infty$ we have, for $j = 1, \ldots, K$, $N_{j}(M)\rightarrow\infty$ in a way that $N \neq M$ and $M/N_{j}
\rightarrow c_{j}$ for some constant $0<c_{j}<1$.

Under the above conditions, it follows from \cite{roberto_icaspp2023consistent}, that $d_M - \hat{d}_M \to 0$ with probability one, where the consistent estimator $\hat{d}_M$ can be expressed as follows. 
Let us denote the eigenvalues and the associated eigenvectors of the SCM $\hat{\mathbf{R}}_j$ as $\hat{\lambda}^{(j)}_1 < \ldots < \hat{\lambda}^{(j)}_M$ and $\hat{\mathbf{e}}^{(j)}_1, \ldots, \hat{\mathbf{e}}^{(j)}_M$, respectively. The consistent estimator $\hat{d}_{M}$ can be expressed as
\[
\hat{d}_{M}=\alpha^{(1)}+\alpha^{(2)}-\frac{2}{M}\sum_{k=1}^{M}\sum_{m=1}^{M}\beta_{k}^{(1)}\beta_{m}^{(2)}\left\vert \left(  \hat{\mathbf{e}}_{k}^{(1)}\right)  ^{H}\hat{\mathbf{e}}_{m}^{(2)}\right\vert ^{2}
\]
where (for  $j = 1,\ldots, K$) the coefficients $\beta_{k}^{(j)}$ and $\alpha_{k}^{(j)}$, $k=1,\ldots,M$, are defined as
\begin{align*}
 \beta_{k}^{(j)}  & =\left(  1+\sum_{\substack{m=1\\m\neq k}}^{M}\frac{\hat{\lambda}_{k}^{(j)}}{\hat{\lambda}_{m}^{(j)}-\hat{\lambda}_{k}^{(j)}}-\sum_{m=1}^{M}\frac{\hat{\mu}_{k}^{(j)}}{\hat{\lambda}_{m}^{(j)}-\hat{\mu}_{k}^{(j)}}\right)  \log\hat{\lambda}_{k}^{(j)}
\\&
\sum_{\substack{r=1\\r\neq k}}^{M}\frac{\hat{\lambda}_{r}^{(j)}}{\hat{\lambda}_{r}^{(j)}-\hat{\lambda}_{k}^{(j)}}\log\hat{\lambda}_{r}^{(j)}-\sum_{r=1}^{M}\frac{\hat{\mu}_{r}^{(j)}}{\hat{\mu}_{r}^{(j)}-\hat{\lambda}_{k}^{(j)}}\log\hat{\mu}_{r}^{(j)} + 1
\end{align*}
and
\begin{multline}
\alpha^{(j)}=
\left(  \frac{N_{j}}{M}-1\right)  \sum_{r=1}^{M}\left(1+\log\hat{\mu}_{r}^{(j)}\right)  ^{2}-\left(  1+\log\hat{\lambda}_{r}^{(j)}\right)^{2}
\\
+
\frac{1}{M}\sum_{k=1}^{M}\left(  1+\log\hat{\lambda}_{k}^{(j)}\right)^{2}{-}\left(  \frac{N_{j}}{M}-1\right)  \log^{2}\left(  1-\frac{M}{N_{j}}\right)  + 1
\\
+\frac{2}{M}\sum_{k=1}^{M}\sum_{r=1}^{M}\left[  \Phi_{2}\left(  \frac{\hat{\mu}_{r}^{(j)}}{\hat{\lambda}_{k}^{(j)}}\right)  -\Phi_{2}\left(\frac{\hat{\lambda}_{r}^{(j)}}{\hat{\lambda}_{k}^{(j)}}\right)  \right] 
 \\
+\frac{2}{M}\sum_{k=1}^{M}\Biggl(  \sum_{\substack{r=1\\r\neq k}}^{M}\log\frac{\hat{\lambda}_{r}^{(j)}}{\hat{\lambda}_{k}^{(j)}}\log\frac{\hat{\lambda}_{k}^{(j)}}{\left\vert \hat{\lambda}_{k}^{(j)}-\hat{\lambda}_{r}^{(j)}\right\vert } 
\\ 
-\sum_{r=1}^{M}\log\frac{\hat{\mu}_{r}^{(j)}}{\hat{\lambda}_{k}^{(j)}}\log\frac{\hat{\lambda}_{k}^{(j)}}{\left\vert \hat{\lambda}_{k}^{(j)}-\hat{\mu}_{r}^{(j)}\right\vert }\Biggr) 
\label{eq:definition_alpha_j}
\end{multline}
respectively. In the above equation, we have denoted by $\hat{\mu}_{1}^{(j)}<\ldots<\hat{\mu}_{M}^{(j)}$ the $\bar{M}_j$
solutions to 
$
 1 = {N_j}^{-1} \mathrm{tr}[\hat{\mathbf{R}}_j\hat{\mathbf{Q}}_j(\hat{\mu}^{(j)})]
$
and we have introduced the function 
 \begin{equation}
\Phi_{2}(x) 
=\left\{
\begin{array}
[c]{ccc}
\mathrm{Li}_{2}\left(  x\right)  &  & x<1\\
\frac{\pi^{2}}{3}-\frac{1}{2}\log^{2}x-\mathrm{Li}_{2}\left(  x^{-1}\right)  &
& x\geq1
\end{array}
\right.  \label{eq:definitionPhi(x)Li2}
\end{equation}
where $\mathrm{Li}_{2}\left(  x\right)  =-\int_{0}^{x} y^{-1}\log(1-y) dy$ is the di-logarithm.

 \section{Asymptotic Fluctuation of $\hat{d}_M$}
\label{sec:main_result}

The results in the previous section were derived for the comparison of two SCMs. In practical scenarios, however, one is usually interested in considering multiple distances among several covariances and how they relate to each other. 
In this context, let $\hat{\mathbf{d}}_M$ denote an $R$-dimensional column vector containing the consistent estimator of the log-Euclidean distances between $R$ different pairs from a total of $K$ transmitters, that is 
\begin{equation} 
\label{eq:defDistvect}
\hat{\mathbf{d}}_M = \left[ \hat{d}_M(i_1,j_1), \ldots \hat{d}_M(i_R,j_R) \right]^T
\end{equation}
where 
 $1  \leq i_r, j_r \leq K, i_r \neq j_r$, $r=1,\ldots,R$ 
is the distance between 
two different UEs (indexed by $i_r$ and $j_r$) among the $R$ considered pairs. 
Likewise, we will denote by $\mathbf{d}_{M}$ the $R$-dimensional column vector containing the actual distances between these covariances, so that (by the above discussion), $|\hat{\mathbf{d}}_M - \mathbf{d}_M| \to \mathbf{0}_M$ almost surely. In this section, we will show that the random vector  $\boldsymbol{\zeta}_M =  M(  \hat{\mathbf{d}}_{M}-\mathbf{d}_{M})$ of size $R \times 1$ asymptotically fluctuates according to a multivariate Gaussian distribution with zero mean and certain covariance.  

Let us introduce the expression of the asymptotic covariance matrix here. To that effect, let us consider the resolvent  matrix ${\mathbf{Q}}_{j}(\omega)  = (  {\mathbf{R}}_{j}-\omega\mathbf{I}_{M} )^{-1}$. The asymptotic covariance matrix $\bar{\boldsymbol{\Sigma}}_M$ is defined as an $R \times R$ matrix with $(r,s)$th entry given by 
\begin{multline}
    \left[\bar{\boldsymbol{\Sigma}}_M\right]_{r,s} = \frac{1}{(2\pi\mathrm{j})^4} \oint\nolimits_{\mathrm{C}_{\omega_{i_r}}}\oint\nolimits_{\mathrm{C}_{\omega_{j_r}}}\oint\nolimits_{\mathrm{C}_{\omega_{i_s}}}\oint\nolimits_{\mathrm{C}_{\omega_{j_s}}}
    \\
     \bigl(\log(\omega_{i_r})
     -
     \log(\omega_{j_r})\bigr)^2 
     \bigl(\log(\tilde{\omega}_{i_s})
     -
     \log(\tilde{\omega}_{j_s})\bigr)^2 \times
     \\
    \times \bar{\sigma}^2_{i_r,j_r,i_s,j_s}(\omega_{i_r},\omega_{j_r},\tilde{\omega}_{i_s},\tilde {\omega}_{j_s}) d\omega_{i_r} d\omega_{j_r} d\tilde{\omega}_{i_s} d\tilde{\omega}_{j_s}
\end{multline}
where $\mathrm{C}_{\omega_{j}}$ is a simple contour enclosing all the eigenvalues of $\mathbf{R}_j$ and not $\{0\}$, and where $\bar{\sigma}^2_{i,j,m,n}(\omega_{i},\omega_{j},\tilde{\omega}_{m},\tilde{\omega}_{n}) $ is defined as
\begin{align*}
\bar{\sigma}^{2}_{i,j,m,n}& \left(  \omega_{i},\omega_{j},  \tilde{\omega}_{m},\tilde{\omega}_{n}\right) 
 = {\varrho}_{i,j}(\omega_i,\omega_j,\tilde{\omega}_m,\tilde{\omega}_n) \delta_{i=m}\delta_{j=n} 
\nonumber\\
&+ \varrho_{i,j}(\omega_i,\omega_j,\tilde{\omega}_n,\tilde{\omega}_m) \delta_{i=n}\delta_{j=m}
\\
& + \bar{\sigma}_{i}^{2}\left(  \omega_{i},\tilde{\omega}_{m};\mathbf{Q}_{j}\left(
\omega_{j}\right)  ,\mathbf{Q}_{n}(\tilde{\omega}_{n})\right) \delta_{i=m}    
\nonumber\\
&+ \bar{\sigma}_{j}^{2}\left(  \omega_{j},\tilde{\omega}_{n};\mathbf{Q}_{i}\left(
\omega_{i}\right)  ,\mathbf{Q}_{m}(\tilde{\omega}_{m})\right) \delta_{j=n}  \nonumber \\
& + \bar{\sigma}_{i}^{2}\left(  \omega_{i},\tilde{\omega}_{n};\mathbf{Q}_{j}\left(
\omega_{j}\right)  ,\mathbf{Q}_{m}(\tilde{\omega}_{m})\right) \delta_{i=n}  
\nonumber\\
&+ \bar{\sigma}_{j}^{2}\left(  \omega_{j},\tilde{\omega}_{m};\mathbf{Q}_{i}\left(
\omega_{i}\right)  ,\mathbf{Q}_{n}(\tilde{\omega}_{n})\right) \delta_{j=m}  \nonumber \\ 
\end{align*}
with the following additional definitions. The functions $\bar{\sigma}_{j}^{2}\left(  \omega,\tilde{\omega};\mathbf{A},\mathbf{B}\right) $ for $j = 1,\ldots, K$ are defined as
 \begin{multline*}
\bar{\sigma}_{j}^{2}\left(  \omega,\tilde{\omega};\mathbf{A},\mathbf{B}\right)   
=
\frac{\mathrm{tr}
\left[  \tilde{\Gamma}_{j}(\omega,\tilde{\omega}) \mathbf{A}
 \tilde{\Gamma}_{j}(\omega,\tilde{\omega}) \mathbf{B}  \right] }{N_j  \left(1-\Gamma_{j} (\omega, \tilde{\omega}) \right)} \\
 +\frac{\mathrm{tr}\left[  \mathbf{R}_{j}\mathbf{Q}_{j}\left(
\omega\right)  \tilde{\Gamma}_{j} (\omega,\tilde{\omega})\mathbf{B} \right]
\mathrm{tr}\left[  \mathbf{R}_{j}\mathbf{Q}_{j}\left(
\omega\right)  \tilde{\Gamma}_{j} (\omega,\tilde{\omega}) \mathbf{A} \right]
} {N_j^2\left( 
1-\Gamma_{j}\left(\omega, \Tilde{\omega} \right)\right)^{2} }
 \end{multline*}
and where we have introduced the quantities
\begin{align*}
\Gamma_{j}(\omega,\tilde{\omega}) &=\frac{1}{N_{j}}\mathrm{tr}\left[
\mathbf{R}_{j}^{2} \mathbf{Q}_j\left(\omega\right) \mathbf{Q}_j\left(\tilde{\omega}\right)
\right]
\\
\tilde{\Gamma}_{j}(\omega,\tilde{\omega})  
&=
\mathbf{R}_{j} \mathbf{Q}_j\left(\omega\right) \mathbf{Q}_j\left(\tilde{\omega}\right)
.
\end{align*}
Finally, the functions ${\varrho}_{i,j}(\omega_i,\omega_j,\tilde{\omega}_i,\tilde{\omega}_j) $ are defined as 
\begin{multline}
    {\varrho}_{i,j}(\omega_i,\omega_j,\tilde{\omega}_i,\tilde{\omega}_j) =
    \\
\frac{\mathrm{tr}^{2}\left[  \mathbf{R}_{i}\mathbf{Q}_{i}\left(  \omega
_{i}\right)  \mathbf{Q}_{i}\left( \tilde{\omega}_{i}\right)  \mathbf{R}
_{j}\mathbf{Q}_{j}\left(  \omega_{j}\right)  \mathbf{Q}_{j}\left(  \tilde{\omega}_{j}\right)  \right]  }{N_{i}N_{j}\left(  1-\Gamma_{i}(\omega
_{i},\tilde{\omega}_{i})\right)  \left(  1-\Gamma_{j}(\omega_{j},\tilde{\omega}
_{j})\right)  }.  
\end{multline}

Having introduced the terms above, we are now in the position to formulate a central limit theorem. 
\begin{theorem} 
\label{th:cltEstimators}
In addition to \textbf{(As1)}-\textbf{(As2)}, assume that the observations are complex circularly symmetric
Gaussian distributed and that the minimum eigenvalue of $\bar{\boldsymbol{\Sigma}}_M$ is bounded away from zero. Then, the random vector 
\[
\bar{\boldsymbol{\Sigma}}_M^{-1/2}\left[M(\hat{\mathbf{d}}_M - \mathbf{d}_M)\right] 
\]
converges in law to a multivariate standard Gaussian. 
\end{theorem}
\begin{IEEEproof}
    See \cite{pereira23tit}.
\end{IEEEproof}
Even if the expression obtained above appears to be difficult to evaluate due to the presence of the contour integrals, one can  typically simplify these expressions using conventional residue calculus (more details are given in  \cite{pereira23tit}).

More importantly, the above theorem describes the asymptotic behavior of the $R-$dimensional vector $\hat{\mathbf{d}}_M$. This, in turn, allows to infer important statistical insights regarding the quality of the clustering of different SCMs. In what follows, we will explore its practical application to the wireless communications context described above.



\section{Numerical Validation}

\begin{figure}[b]
    \centering
    \includegraphics[width=0.4\linewidth]
    {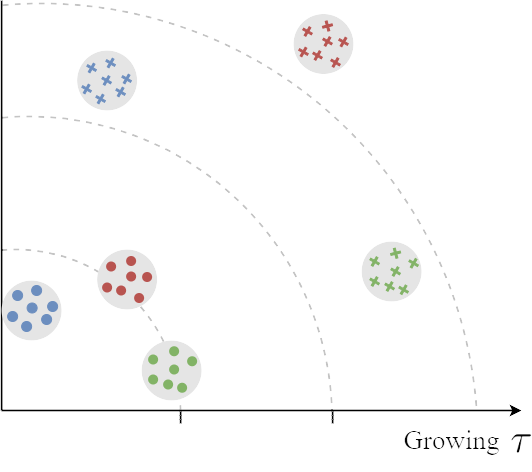}
    \caption{Illustrative example of the simulation scenario with $G=3$ groups (blue, red, and green), $K=21$ UEs 
    and two different values of $\tau= 1,2$ (represented by the circles and the cross). From the BS perspective (located at the origin), the parameter $\tau$ controls the cluster separability.   
    }
    \label{fig:illustrative_scenario}
\end{figure}

In order to validate the results presented above, we consider the scenario where $K$ distinct transmitters are each uniquely assigned to one of three groups ($g = 1, 2, 3$) based on their spatial locations. More specifically, we fix three cluster centroids with coordinates (in meters) at $(x, y) = {(20, 60)\tau, (70, 70)\tau, (80, 30)}\tau$, where $\tau>0$ is a certain scaling parameter that effectively controls how close these clusters are.
Each transmitter within a group is randomly placed within a
$15$ meter
radius from its group's centroid location, while ensuring that no two transmitters are placed at the same location. Figure~\ref{fig:illustrative_scenario} illustrates the general setup considered in this work. 
 for two different values of $\tau$. Notice that, from the BS perspective, small values of $\tau$  result in the different groups being close to each other, whereas larger values lead to groups being further away from each other.  

We follow a similar model as the one detailed in \cite{emuil2021cellfree_book} used to model 3GPP channels, e.g. the Urban Microcell model. 
Hence, we model the large-scale fading coefficient, measured in decibels dB for each transmitter, as 
$$
\beta_j = -30.5 - 36.7 \log_{10} (d_j) + \varsigma_j
$$
where $d_j, j = 1, \ldots, K$  is the three-dimensional physical distance (in meters) between the $j$th transmitter to the BS. The shadowing is designed as a random variable with zero mean and is correlated among different users by $\mathbb{E}[\varsigma_j \varsigma_j] = 4^2 2^{-1\delta_{jk}}$, where $\delta_{jk}$ is the distance between the $j$th and $k$th UE (see \cite[Chapter 5]{emuil2021cellfree_book} for details on the fading modeling). 
Additionally, we consider an uplink communication scenario utilizing a  $20$MHz bandwidth, with the total receiver noise power of $-94$dBm. We assume the BS to be located at $(0,0)$ and to be elevated $12$ meters above  the ground, whereas all the transmitters are placed $2$ meters above from the ground.

Finally, we consider non-line-of-sight communication only and model the channel's covariance matrices $\mathbf{R}_j, j = 1\ldots, K$ as 
the contribution of multi-path signals impinging over a $M$ dimensional uniform linear array, namely
$$
\mathbf{R}_j  = \beta_j 
\int_{-\pi}^{\pi} \mathbf{a}(\theta)\mathbf{a}^H(\theta) g_j(\theta)d\theta
$$
where $\mathbf{a}(\theta)$ is the array response at the base station's antenna array at angle $\theta$ and $g_j(\theta)$ represents the probability density function of a Gaussian distribution with mean $\bar{\theta}_j=\arctan(y_j / x_j)$ (i.e., depending on the user's location) and variance $30^\circ$ (fixed for all transmitters).
We assume for simplicity of exposition that the number of channel samples $N^{(g)}, g = 1, 2, 3$~are the same among all the UEs associated to the same cluster centroid, so that we will denote $c^{(1)} = M/N^{(1)}, c^{(2)} = M/N^{(2)}, c^{(3)} = M/N^{(3)}$ to represent the ratio between the number of antennas at the BS and the number of samples at groups $g = 1,2,3$.


We start by comparing  the asymptotic distribution described in Theorem~\ref{th:cltEstimators} with the empirical distribution of the estimated distances with finite dimensions. In order to simplify exposition, Figure~\ref{fig:illustrative_scenario} compares the histogram (in blue) and the asymptotic distributions (in orange) of these metrics for a pair of UEs (i.e., $R = 1$) with $\beta_1 = \beta_2 = 1$ and $N_1 = N_2 = 24$ for some specific choices of  $M, \theta_1, \theta_2 $.
Observe that there is a very good match between the asymptotic and empirical distributions over the different scenarios, which further highlights the accuracy of our result. 
Specifically, a good match between the asymptotic mean and empirical expectation  suggests that the consistent estimator accurately approximates the true distance (i.e., $d_M - \hat{d}_M \to 0$), while a good alignment between asymptotic and empirical variance corroborates  for the results presented in Theorem~\ref{th:cltEstimators}. 
Additionally, we observe that despite Theorem~\ref{th:cltEstimators} being designed considering the large regime assumptions, results seems to be consistent even for relatively low system dimensions. 

\begin{figure}[t]
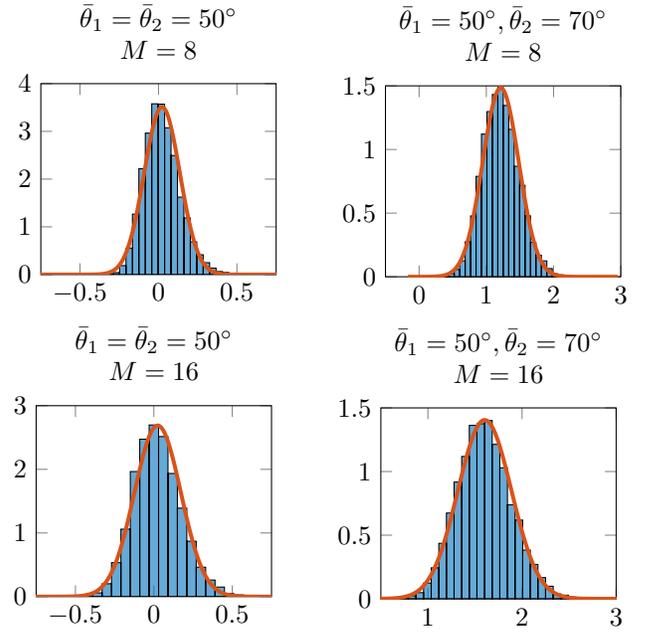

     \centering
 \begin{subfigure}[b]{0.23\textwidth}
         \centering
        {\include{fig/clt/M_8_24_24_same}}
              \vspace*{-0.3in} 
 \end{subfigure}
     \hspace*{0.05in}
 \begin{subfigure}[b]{0.23\textwidth}
         \centering
        {\include{fig/clt/M_8_24_24_diff}}
              \vspace*{-0.3in} 
 \end{subfigure}
 \begin{subfigure}[b]{0.23\textwidth}
         \centering
        {\include{fig/clt/M_16_24_24_same}}
 \end{subfigure}
     \hspace*{0.05in}
 \begin{subfigure}[b]{0.23\textwidth}
         \centering
        {\include{fig/clt/M_16_24_24_diff}}
 \end{subfigure}
      \vspace{-1.5\baselineskip}
      \caption{Comparison between empirical distribution of $\hat{d}_M$ and its asymptotic according to Theorem \ref{th:cltEstimators} under different scenarios and $N_1 = N_2 = 24$.
      }
    \label{fig:histogram} 
    \vspace{-1\baselineskip}
\end{figure}

Finally, note in the cases where $\bar{\theta}_1 = \bar{\theta}_2$ (i.e., implying identical covariance matrices), the consistent estimator  closely approximates zero. Conversely,  when $\bar{\theta}_1 \neq \bar{\theta}_2$, the values of $\hat{d}_M$ become significantly greater than zero. This distinction becomes particularly useful when clustering several sample covariance matrices (SCMs), enabling more accurate differentiation between distinct groups.


\subsection{Clustering of Multiple UEs}

We now consider the clustering of multiple SCMs in the wireless communications context. Specifically, 
we explore the advantages of using the consistent estimate over the traditional \textit{plug-in} estimator to cluster multiple UEs based on their SCMs. In this context, we examine the wireless simulation scenario outlined above  and illustrated in Figure~\ref{fig:illustrative_scenario}, with three groups  ($G = 3$) and different values of $K$. 
For clarity, in the figures, the traditional \textit{plug-in} estimator is denoted as ``TRAD'', while the consistent estimator is labeled as ``IMP''.

We consider that the clustering using a specific distance is successful for a specific set of channel realizations when the inter-cluster distance is always higher than the largest intra-cluster distance. We can therefore evaluate the probability of clustering success by conditioning on each of the three clusters achieving the maximum intra-cluster distance. Let $\mathcal{J}_\mathrm{inter}$ denote the set of pairs of indexes of UEs belonging to different clusters and $\mathcal{J}_\mathrm{intra} (g)$ the set of pairs of indexes of UEs belonging to the $g$ cluster. The probability of success can be mathematically expressed as 
\begin{multline*}
\mathbb{P}_\mathrm{succ}  =  \sum_{g=1}^{3} \mathbb{P} \Bigg[  \bigcap_{(i,j)\in \mathcal{J}_\mathrm{inter}} \bigg\{ \hat{d}_{M} (i,j) > \\ 
>\max_{(k,l) \in \mathcal{J}_\mathrm{intra}(g)} \hat{d}_M(k,l) \bigg\} \Bigg\vert \mathcal{A}_g \Bigg]
\mathbb{P} \left[\mathcal{A}_g \right]
\end{multline*}
where we have denoted as $ \mathcal{A}_g$ the event that the $g$th cluster achieves the maximum intra-cluster distance. 

Each of these  probabilities can in turn be written as
$
\mathbb{P}\left( \mathbf{A}\hat{\mathbf{d}}_M  < 0\right)
$
where $\mathbf{d}_M$ is a column vector that contains all the distances as in (\ref{eq:defDistvect}) and $\mathbf{A}$ is a selection matrix with all the entries equal to zero except for one $+1$ and one $-1$ for each row, corresponding to the selection of distances to conform the different events that come into play into each argument of the above expression. Since $\hat{\mathbf{d}}_M$ is asymptotically Gaussian distributed, so is the resulting column vector $\mathbf{A}\hat{\mathbf{d}}_M$. In fact, the transformed vector will also be asymptotically approximated as Gaussian distributed, with mean $\mathbf{A}\mathbf{d}_M$ and covariance $\mathbf{A}(\bar{\mathbf{\Sigma}}_M / M^2)\mathbf{A}^\mathrm{T} $. Hence, each of the probabilities above can be evaluated by a multivariate Gaussian cumulative distribution function evaluated at zero.  
To evaluate the empirical probability of accurate clustering (referred to as ``Prob.  Clustering'' in the figures), we will utilize $10^3$ simulations. This clustering success rate is then defined as the percentage of realizations where the all estimated intra-cluster distances are lower than the smallest estimated inter-cluster distance.

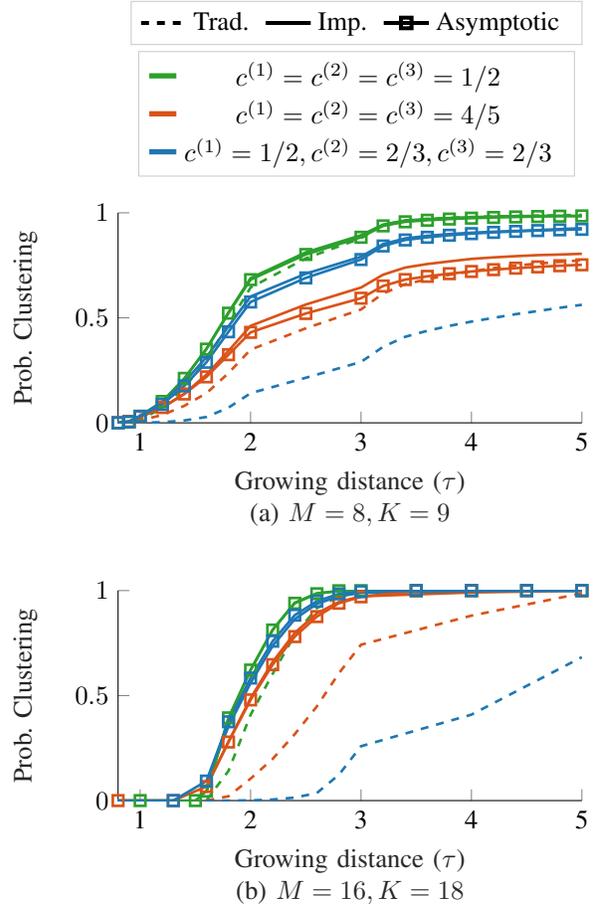
\begin{figure}[tb]
\hspace{-0.25in}
\begin{subfigure}{0.9\textwidth}
{\vspace{-2.5em}\input{fig/fig3_dist/legend_dist_style}\vspace{-2.5em}}
\end{subfigure}
\hspace*{-0.25in}
\begin{subfigure}{0.9\textwidth}
{\input{fig/fig3_dist/legend_dist}}
\hspace*{-10in} 
\end{subfigure}
     \flushleft 
     \vspace*{-0.15in} 
 \begin{subfigure}[b]{0.4\textwidth}
         \flushleft 
        {\include{fig/fig3_dist/fig3_a_less} }
              \vspace*{-0.3in} 
 \end{subfigure}
 \begin{subfigure}[b]{0.4\textwidth}
         \flushleft 
        {\include{fig/fig3_dist/M_16_kg_6}}
              \vspace*{-0.3in} 
 \end{subfigure}
      \caption{Probability of correct clustering (y-axis) $K$ UEs into three groups,
       different ratios $M/N^{(g)}, g=1,2,3$ and growing $\tau$ (x-axis).
      }
    \label{fig:result:systemgrowth} 
    \vspace{-1\baselineskip}
\end{figure}

The discussion above allow us to asymptotically study the behavior of the log-Euclidean metric when employed to the clustering of SCMs of different UEs. 
Note that this is only possible due to Theorem~\ref{th:cltEstimators} which, as shown above, allows us to properly approximate the asymptotic behavior of the log-Euclidean metric.  Figure~\ref{fig:result:systemgrowth} portrays the probability of correct clustering for the traditional log-Euclidean estimator (dashed lines), the consistent estimator (solid lines) and its asymptotic equivalent (solid lines with square marker) for different regimes $M/N^{(1)}, M/N^{(2)}, M/N^{(3)}$ and growing $\tau$. 
Note that there exists a very good alignment between the empirical and theoretical probabilities of correctly clustering the different SCMs, indicating the correctness of our analytical results and their efficacy as prediction mechanisms.  This alignment happens regardless of the scenario (different values of $c^{(g)}, g=1, 2, 3$) and number of UEs considered. 
Additionally, as illustrated above, larger values of $\tau$ reflects in easier to cluster scenarios.
Nonetheless, the consistent estimator consistently outperforms the tradition \textit{plug-in} and  produces a high probability of clustering even for relatively small values of $\tau$ and in scenarios where the \textit{plug-in} underperforms. This  indicates the robustness  of the consistent estimator under varying system conditions.

\begin{figure}[tb]
\hspace{-0.2in}
\begin{subfigure}{0.9\textwidth}
    {\vspace{-2em}\hspace{-5em}\input{fig/fig4_fixedM/legend_fig4}}
\end{subfigure}
\hspace{-0.05in}
 \begin{subfigure}[b]{0.4\textwidth}
         \centering
{\include{fig/fig4_fixedM/fig4_m8}}
              \vspace*{-0.3in} 
 \end{subfigure}

      \caption{
      Probability of correct clustering (y-axis) $K$ UEs into three groups for fixed $M = 8$, $\tau=1.5$ and growing $N^{(g)}$.
      }
    \label{fig:result:growing_N_fixedM} 
    \vspace{-1\baselineskip}
\end{figure}
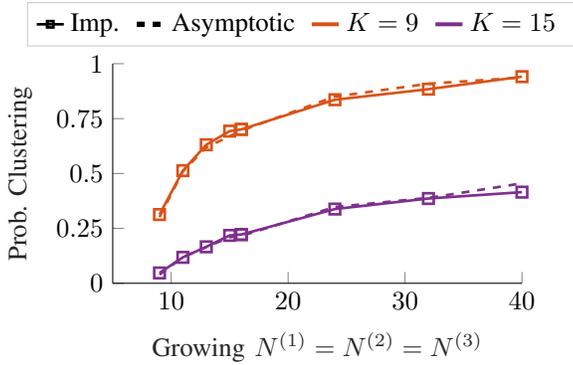

Our results also allow us to predict the necessary number of samples to reach a certain accuracy level.  
Moreover, in the context of wireless communications, the number of samples strongly affect the quality of the clustering solution.
However,  performing a large number of channel estimations can be prohibitively expensive or even impractical, depending on the scenario.
Consequently, it is often advantageous,
from both energy efficiency and system design perspectives,
to decide a priori the number of channel estimations required by the system to reach a certain clustering quality. Therefore, in Figure~\ref{fig:result:growing_N_fixedM}, we relax the assumption \textbf{(As2)} and consider the scenario where $M$ is fixed and $N^{(g)}, g=1,2,3$ grows unbounded
. Notice that, despite relaxing this assumption, our asymptotic prediction mechanism remains closely aligned with the empirical results, demonstrating the robustness of our approach.


\section{Conclusion}

In this work, we have proposed a statistical framework to estimate the probability of correct clustering multiple UEs via their channels' sample covariance matrices using a consistent estimator of the log-Euclidean distance of two SCMs.
Numerical simulations confirm that the consistent estimator can increase the quality of clustering solutions even for relative low number of channel samples. Finally, we show that the proposed framework not only enhances the  clustering accuracy but also provides a reliable predictive methodology for assessing the performance of these clustering algorithms in realistic MU-MIMO settings.

\bibliographystyle{./bibliography/IEEEtran}
\bibliography{./bibliography/IEEEexample}

\end{document}

%% file: fig/fig3_dist/legend_dist_style.tex
\definecolor{mycolor1}{RGB}{44,160,44}
\definecolor{mycolor2}{rgb}{0.85000,0.32500,0.09800}%
\definecolor{mycolor3}{RGB}{31,119,180}

\definecolor{mycolor4}{rgb}{0.49400,0.18400,0.55600}%
\definecolor{mycolor5}{RGB}{31,119,180}%

\definecolor{darkgray176}{RGB}{176,176,176}
\definecolor{darkorange25512714}{RGB}{255,127,14}
\definecolor{forestgreen4416044}{RGB}{44,160,44}
\definecolor{lightgray204}{RGB}{204,204,204}
\definecolor{magenta}{RGB}{255,0,255}
\definecolor{orange2551868}{RGB}{255,186,8}
\definecolor{saddlebrown164660}{RGB}{164,66,0}
\definecolor{steelblue31119180}{RGB}{31,119,180}
\definecolor{saddlebrown164660}{RGB}{164,66,0}

\begin{tikzpicture} 
\begin{axis}[
hide axis,
width=0.8\linewidth,
height=1in,
at={(-1,0)},
xmin=0,
xmax=0,
ymin=0,
ymax=0.4,
legend style={fill opacity=1, draw opacity=1, text opacity=1, draw=lightgray204, legend columns=3,
    anchor=north,
    at={(-0.2in, 0in)},
}
]

\addlegendimage{black, line width=1.5pt, dashed};
\addlegendentry{Trad. \hspace{1em}};

\addlegendimage{black, line width=1.5pt};
\addlegendentry{Imp. \hspace{1em}};

\addlegendimage{black, line width=1.5pt, mark=square};
\addlegendentry{Asymptotic \hspace{1em}};

\end{axis}

\end{tikzpicture}

%% file: fig/fig3_dist/legend_dist.tex

\definecolor{mycolor4}{rgb}{0.49400,0.18400,0.55600}%


\definecolor{darkgray176}{RGB}{176,176,176}
\definecolor{darkorange25512714}{RGB}{255,127,14}
\definecolor{forestgreen4416044}{RGB}{44,160,44}
\definecolor{lightgray204}{RGB}{204,204,204}
\definecolor{magenta}{RGB}{255,0,255}
\definecolor{orange2551868}{RGB}{255,186,8}
\definecolor{saddlebrown164660}{RGB}{164,66,0}
\definecolor{steelblue31119180}{RGB}{31,119,180}
\definecolor{saddlebrown164660}{RGB}{164,66,0}

\definecolor{mycolor1}{RGB}{44,160,44}
\definecolor{mycolor2}{rgb}{0.85000,0.32500,0.09800}%
\definecolor{mycolor3}{RGB}{31,119,180}

\begin{tikzpicture} 
\begin{axis}[
hide axis,
width=0.8\linewidth,
height=1in,
at={(-1,0)},
xmin=0,
xmax=0,
ymin=0,
ymax=0.4,
legend style={fill opacity=1, draw opacity=1, text opacity=1, draw=lightgray204, legend columns=1,
    anchor=north,
    at={(-0.2in, 0in)},
    legend image post style={scale=0.6}
}
]

\addlegendimage{mycolor1, line width=2pt};
\addlegendentry{$c^{(1)} = c^{(2)}=  c^{(3)} = 1/2$ \hspace{1em}};

\addlegendimage{mycolor2, line width=2pt};
\addlegendentry{$c^{(1)} =  c^{(2)}=  c^{(3)} = 4/5$ \hspace{1em}};


\addlegendimage{mycolor3, line width=2pt};
\addlegendentry{$c^{(1)} = 1/2, c^{(2)}= 2/3, c^{(3)} = 2/3$ \hspace{1em}};

\coordinate (legend) at (axis description cs:-0.0,0.03);

\end{axis}

\end{tikzpicture}

%% file: fig/fig3_dist/fig3_a_less.tex
%
%

\definecolor{mycolor1}{RGB}{44,160,44}
\definecolor{mycolor2}{rgb}{0.85000,0.32500,0.09800}%
\definecolor{mycolor3}{RGB}{31,119,180}

\definecolor{mycolor4}{rgb}{0.49400,0.18400,0.55600}%
\definecolor{mycolor5}{RGB}{31,119,180}%

\begin{tikzpicture}

\begin{axis}[%
width=0.85\linewidth,
height=1.1in,
at={(0in,2.1in)},
scale only axis,
scale only axis,
align =center,
ytick={0, 0.5, 1},
xtick={10,20,30,40,50},
xticklabels={1,2,3,4,5},
xmin=8,
xmax=50,
xlabel style={font=\color{white!15!black}},
xlabel={{Growing distance ($\tau$)} \\ (a) $M=8, K=9$},
ymin=0,
ymax=1,
ylabel style={font=\color{white!15!black}},
ylabel={Prob.  Clustering},
axis background/.style={fill=white},
axis x line*=bottom,
axis y line*=left,
legend style={legend cell align=left, align=left, draw=white!15!black}
]
\addplot [color=mycolor1, dashed, line width=1.0pt]
  table[row sep=crcr]{%
1	0\\
2	0\\
3	0\\
4	0\\
5	0\\
6	0\\
7	0\\
8	0\\
9	0.001\\
10	0.0164\\
12	0.069\\
14	0.1672\\
16	0.2982\\
18	0.4708\\
20	0.6444\\
25	0.7832\\
30	0.8788\\
32	0.9424\\
34	0.9642\\
36	0.9714\\
38	0.9766\\
40	0.98\\
42	0.9828\\
44	0.985\\
46	0.9868\\
48	0.988333333333333\\
50	0.99\\
};

\addplot [color=mycolor1, line width=1.0pt]
  table[row sep=crcr]{%
1	0\\
2	0\\
3	0\\
4	0\\
5	0\\
6	0\\
7	0\\
8	0\\
9	0.0028\\
10	0.0284\\
12	0.0974\\
14	0.2092\\
16	0.3528\\
18	0.5262\\
20	0.6894\\
25	0.8118\\
30	0.8934\\
32	0.9442\\
34	0.9634\\
36	0.9706\\
38	0.9754\\
40	0.9792\\
42	0.9822\\
44	0.9842\\
46	0.985\\
48	0.986\\
50	0.984\\
};

\addplot [color=mycolor1, line width=1.0pt, mark=square, mark options={solid, mycolor1}]
  table[row sep=crcr]{%
1	0\\
2	0\\
3	1.19624821404517e-92\\
4	4.65639018079434e-53\\
5	5.93990640967262e-30\\
6	3.59849016537498e-14\\
7	3.31956190627346e-08\\
8	1.41045947716765e-05\\
9	0.00357009028543197\\
10	0.0314797196675464\\
12	0.101747784027033\\
14	0.211852669673283\\
16	0.350872178353366\\
18	0.522220526599531\\
20	0.681869932795362\\
25	0.801726779340869\\
30	0.88357860109849\\
32	0.937855162327187\\
34	0.957278499140111\\
36	0.964819162228612\\
38	0.970390861359774\\
40	0.974596920405461\\
42	0.977820206538797\\
44	0.98031754720407\\
46	0.982281568643789\\
48	0.984036931604095\\
50	0.985318600530829\\
};

\addplot [color=mycolor2, dashed, line width=1.0pt]
  table[row sep=crcr]{%
1	0\\
2	0\\
3	0\\
4	0\\
5	0\\
6	0\\
7	0\\
8	0\\
9	0.002\\
10	0.0116\\
12	0.0372\\
14	0.081\\
16	0.141\\
18	0.237\\
20	0.3486\\
25	0.4502\\
30	0.5398\\
32	0.6162\\
34	0.6598\\
36	0.6838\\
38	0.7046\\
40	0.7214\\
42	0.7368\\
44	0.7486\\
46	0.7582\\
48	0.766666666666667\\
50	0.774\\
};

\addplot [color=mycolor2, line width=1.0pt]
  table[row sep=crcr]{%
1	0\\
2	0\\
3	0\\
4	0\\
5	0\\
6	0\\
7	0\\
8	0.0002\\
9	0.0072\\
10	0.0274\\
12	0.0712\\
14	0.1384\\
16	0.2296\\
18	0.3428\\
20	0.4618\\
25	0.5634\\
30	0.6442\\
32	0.7054\\
34	0.7394\\
36	0.7562\\
38	0.7688\\
40	0.7806\\
42	0.7878\\
44	0.7936\\
46	0.7986\\
48	0.801333333333333\\
50	0.805\\
};

\addplot [color=mycolor2, line width=1.0pt, mark=square, mark options={solid, mycolor2}]
  table[row sep=crcr]{%
1	0\\
2	1.3421939294869e-282\\
3	1.36368546737005e-33\\
4	1.15860118867495e-16\\
5	6.80769485532831e-10\\
6	4.03801619769346e-07\\
7	2.47408720474467e-05\\
8	0.000407441459344186\\
9	0.0067881841702589\\
10	0.0293578432585537\\
12	0.0730917080171572\\
14	0.137337184830146\\
16	0.218797874838487\\
18	0.324528153430012\\
20	0.430923222084742\\
25	0.520671677343489\\
30	0.593487105993986\\
32	0.651709135774105\\
34	0.682087737036365\\
36	0.697612585206226\\
38	0.71027641799183\\
40	0.720723564965401\\
42	0.729352048094263\\
44	0.736532271425675\\
46	0.742536195067477\\
48	0.748095416060314\\
50	0.752467501177287\\
};

\addplot [color=mycolor3, dashed, line width=1.0pt]
  table[row sep=crcr]{%
1	0\\
2	0\\
3	0\\
4	0\\
5	0\\
6	0\\
7	0\\
8	0\\
9	0\\
10	0.0004\\
12	0.003\\
14	0.0112\\
16	0.0278\\
18	0.0724\\
20	0.1412\\
25	0.2146\\
30	0.29\\
32	0.362\\
34	0.4098\\
36	0.4382\\
38	0.4618\\
40	0.482\\
42	0.4998\\
44	0.5174\\
46	0.5322\\
48	0.548\\
50	0.562\\
};

\addplot [color=mycolor3, line width=1.0pt]
  table[row sep=crcr]{%
1	0\\
2	0\\
3	0\\
4	0\\
5	0\\
6	0\\
7	0\\
8	0.0002\\
9	0.0064\\
10	0.034\\
12	0.0988\\
14	0.1938\\
16	0.3124\\
18	0.4598\\
20	0.6014\\
25	0.71\\
30	0.7912\\
32	0.8512\\
34	0.8782\\
36	0.8906\\
38	0.899\\
40	0.9052\\
42	0.91\\
44	0.9138\\
46	0.9176\\
48	0.922333333333333\\
50	0.927\\
};

\addplot [color=mycolor3, line width=1.0pt, mark=square, mark options={solid, mycolor3}]
  table[row sep=crcr]{%
1	0\\
2	0\\
3	3.43124641662076e-59\\
4	1.48462165085231e-35\\
5	1.60518751454269e-19\\
6	3.47144672610437e-10\\
7	1.85819111880694e-06\\
8	8.27966088497455e-05\\
9	0.00521924395655911\\
10	0.0311928071716933\\
12	0.088216401129715\\
14	0.175829155466063\\
16	0.288315942091589\\
18	0.433688774889471\\
20	0.575497347426598\\
25	0.690296206219641\\
30	0.777555893938694\\
32	0.842219487149435\\
34	0.870775495097984\\
36	0.883493419858365\\
38	0.893303441848739\\
40	0.900960252589341\\
42	0.906975791407626\\
44	0.911728490781298\\
46	0.915503178580092\\
48	0.918899321419216\\
50	0.921373741557178\\
};

\end{axis}

\end{tikzpicture}%

%% file: fig/fig3_dist/M_16_Kg_6.tex
%
%

\definecolor{mycolor1}{RGB}{44,160,44}
\definecolor{mycolor2}{rgb}{0.85000,0.32500,0.09800}%
\definecolor{mycolor3}{RGB}{31,119,180}

\begin{tikzpicture}

\begin{axis}[%
width=0.85\linewidth,
height=1.1in,
at={(0in,2.1in)},
scale only axis,
scale only axis,
align =center,
ytick={0, 0.5, 1},
xtick={10,20,30,40,50},
xticklabels={1,2,3,4,5},
ytick={0,  0.5,  1},
xmin=8,
xmax=50,
xlabel style={font=\color{white!15!black}},
xlabel={{Growing distance ($\tau$)} \\ (b) $M = 16, K= 18$},
ymin=0,
ymax=1,
ylabel style={font=\color{white!15!black}},
ylabel={Prob.  Clustering},
axis background/.style={fill=white},
axis x line*=bottom,
axis y line*=left,
legend style={legend cell align=left, align=left, draw=white!15!black}
]
\addplot [color=mycolor1, dashed, line width=1.0pt]
  table[row sep=crcr]{%
16	0.001\\
18	0.141666666666667\\
20	0.4086\\
22	0.6036\\
24	0.7894\\
26	0.9166\\
28	0.9736\\
30	0.993\\
40	0.999\\
50	1\\
};

\addplot [color=mycolor1, line width=1.0pt]
  table[row sep=crcr]{%
16	0.038\\
18	0.39\\
20	0.6222\\
22	0.814\\
24	0.9422\\
26	0.9874\\
28	0.9974\\
30	0.9992\\
40	1\\
50	1\\
};

\addplot [color=mycolor2, dashed, line width=1.0pt]
  table[row sep=crcr]{%
16	0\\
18	0.0206666666666667\\
20	0.1044\\
22	0.1986\\
24	0.3164\\
26	0.4484\\
28	0.6066\\
30	0.7422\\
40	0.880666666666667\\
50	0.987\\
};

\addplot [color=mycolor2, line width=1.0pt]
  table[row sep=crcr]{%
16	0.067\\
18	0.281333333333333\\
20	0.4874\\
22	0.6604\\
24	0.7984\\
26	0.8918\\
28	0.9458\\
30	0.9728\\
40	0.991\\
50	1\\
};

\addplot [color=mycolor3, dashed, line width=1.0pt]
  table[row sep=crcr]{%
16	0\\
18	0\\
20	0.001\\
22	0.005\\
24	0.0142\\
26	0.036\\
28	0.1232\\
30	0.2588\\
40	0.409333333333333\\
50	0.683\\
};

\addplot [color=mycolor3, line width=1.0pt]
  table[row sep=crcr]{%
16	0.086\\
18	0.354\\
20	0.5602\\
22	0.7368\\
24	0.8644\\
26	0.937\\
28	0.973\\
30	0.9892\\
40	0.997666666666667\\
50	1\\
};

\addplot [color=mycolor1, line width=1.0pt, mark=square, mark options={solid, mycolor1}, forget plot]
  table[row sep=crcr]{%
0	0.\\
5	0.\\
10	0.\\
15	0.\\
16	0.0455812993987684\\
18	0.393858446289763\\
20	0.621715874173532\\
22	0.812273468180575\\
24	0.93961790726636\\
26	0.984924781319348\\
28	0.999206624866124\\
30	0.999773583158629\\
35 1 \\
40 1 \\
45 1\\
50 1 \\
};
\addplot [color=mycolor2, line width=1.0pt, mark=square, mark options={solid, mycolor2}, forget plot]
  table[row sep=crcr]{%
1	0.\\
6	0.\\
8	0.\\
13	0.\\
16	0.0668824272644888\\
18	0.278815820980697\\
20	0.479544232215738\\
22	0.647413253763656\\
24	0.780909803626052\\
26	0.876122150436299\\
28	0.939779018014966\\
30	0.970252779594335\\
35 1 \\
40 1 \\
45 1\\
50 1 \\
};
\addplot [color=mycolor3, line width=1.0pt, mark=square, mark options={solid, mycolor3}, forget plot]
  table[row sep=crcr]{%
3	0.\\
6	0.\\
13	0.\\
16	0.0930199447567762\\
18	0.375370479774092\\
20	0.583372863871882\\
22	0.758933876292574\\
24	0.8830598472747\\
26	0.948903767372404\\
28	0.984588652274961\\
30	0.9951981856992\\
35 1 \\
40 1 \\
45 1\\
50 1 \\
};

\end{axis}

\end{tikzpicture}%

%% file: fig/fig4_fixedM/legend_fig4.tex
\definecolor{mycolor1}{RGB}{44,160,44}
\definecolor{mycolor2}{rgb}{0.85000,0.32500,0.09800}%
\definecolor{mycolor3}{RGB}{31,119,180}

\definecolor{mycolor4}{rgb}{0.49400,0.18400,0.55600}%
\definecolor{mycolor5}{RGB}{31,119,180}%

\definecolor{darkgray176}{RGB}{176,176,176}
\definecolor{darkorange25512714}{RGB}{255,127,14}
\definecolor{forestgreen4416044}{RGB}{44,160,44}
\definecolor{lightgray204}{RGB}{204,204,204}
\definecolor{magenta}{RGB}{255,0,255}
\definecolor{orange2551868}{RGB}{255,186,8}
\definecolor{saddlebrown164660}{RGB}{164,66,0}
\definecolor{steelblue31119180}{RGB}{31,119,180}
\definecolor{saddlebrown164660}{RGB}{164,66,0}

\begin{tikzpicture} 
\begin{axis}[
hide axis,
width=0.8\linewidth,
height=1in,
at={(0,0)},
xmin=0,
xmax=0,
ymin=0,
ymax=0.4,
legend style={fill opacity=1, draw opacity=1, text opacity=1, draw=lightgray204, legend columns=4,
    anchor=north,
    at={(-0.5in, 0in)},
    legend image post style={scale=0.6}
}
]

\addlegendimage{black, line width=1pt, mark=square};
\addlegendentry{Imp. \hspace{1em}};

\addlegendimage{black, line width=2pt, dashed};
\addlegendentry{Asymptotic \hspace{1em}};

\addlegendimage{color=mycolor2, line width=1.5pt};
\addlegendentry{$K = 9$ \hspace{1em}};

\addlegendimage{color=mycolor4, line width=1.5pt};
\addlegendentry{$K = 15$ \hspace{1em}};

\end{axis}

\end{tikzpicture}

%% file: fig/fig4_fixedM/fig4_m8.tex
%
%
\definecolor{mycolor1}{RGB}{44,160,44}
\definecolor{mycolor2}{rgb}{0.85000,0.32500,0.09800}%
\definecolor{mycolor3}{RGB}{31,119,180}

\definecolor{mycolor4}{rgb}{0.49400,0.18400,0.55600}%
\definecolor{mycolor5}{RGB}{31,119,180}%

\begin{tikzpicture}

\begin{axis}[%
width=0.75\linewidth,
height=1.15in,
at={(0in,2.1in)},
scale only axis,
scale only axis,
align =center,
ytick={0, 0.25, 0.5, 0.75, 1},
xmin=5,
xmax=40,
xlabel style={font=\color{white!15!black}},
xlabel={Growing $N^{(1)} = N^{(2)} = N^{(3)}$},
ymin=0,
ymax=1,
ylabel style={font=\color{white!15!black}},
ylabel={Prob.  Clustering},
axis background/.style={fill=white},
axis x line*=bottom,
axis y line*=left,
legend style={at={(1.in,1.5in)},anchor=north,
    legend columns=2}
]

\addplot [color=mycolor2, line width=1.0pt, mark=square, mark options={solid, mycolor2}]
  table[row sep=crcr]{%
9	0.313\\
11	0.513\\
13	0.631\\
15	0.693\\
16	0.701\\
16	0.701\\
24	0.836\\
32	0.884\\
40	0.941\\
};

\addplot [color=mycolor4, line width=1.0pt, mark=square, mark options={solid, mycolor4}]
  table[row sep=crcr]{%
9	0.047\\
11	0.118\\
13	0.166\\
15	0.218\\
16	0.222\\
16	0.222\\
24	0.338\\
32	0.386\\
40	0.415\\
};

\addplot [color=mycolor2, dashed, line width=1.0pt]
  table[row sep=crcr]{%
9	0.301185647318576\\
11	0.509837504377216\\
13	0.619112404853561\\
15	0.669340847073593\\
16	0.695167900396177\\
16	0.695167900396177\\
24	0.850034657681893\\
32	0.909015080795666\\
40	0.936861200909644\\
};


\addplot [color=mycolor4, dashed, line width=1.0pt]
  table[row sep=crcr]{%
9	0.0394049845680349\\
11	0.119414166974076\\
13	0.164631866614317\\
15	0.209926736321171\\
16	0.215410765870303\\
16	0.215410765870303\\
24	0.347394319503574\\
32	0.38887919555607\\
40	0.455023049484862\\
};

\end{axis}

\end{tikzpicture}%